\documentclass[journal]{IEEEtran}
\usepackage{ifpdf}

\usepackage{cite}
\usepackage{hyperref}

\ifCLASSINFOpdf
\usepackage[pdftex]{graphicx}
\else
\fi
\usepackage{amsmath,amssymb}
\usepackage{array}
\usepackage{threeparttable}
\begin{document}
	%
	\title{Design of a Low Voltage Analog-to-Digital Converter using Voltage Controlled Stochastic Switching of Low Barrier Nanomagnets }
	%
	%
	%
	
	\author{Indranil~Chakraborty, Amogh~Agrawal
		and~Kaushik~Roy,~\IEEEmembership{Fellow,~IEEE}
        \vspace{-6mm}
		\thanks{I~Chakraborty, A~Agrawal and K~Roy are with the Department of Electrical and Computer Engineering, Purdue University, West Lafayette,
			IN, 47907 USA e-mail: ichakra@purdue.edu;agrawa64@purdue.edu;kaushik@purdue.edu}}
	\maketitle
	
	\begin{abstract}
		The inherent stochasticity makes nanoscale devices prospective candidates for low-power computations. Such devices have been demonstrated to exhibit probabilistic switching between two stable states to achieve stochastic behavior. Recently, superparamagnetic nanomagnets (having low energy barrier $E_B \approx 1kT$)  have shown promise of achieving stochastic switching at gigahertz rates, with very low currents. On the other hand, voltage-controlled switching of nanomagnets through the Magneto-electric (ME) effect has shown further improvements in energy efficiency. In this simulation paper, we model a voltage controlled spintronic device based on superparamagnetic nanomagnets exhibiting telegraphic switching characteristics and analyze its behavior under the influence of external bias. Subsequently, we show that the device leverages the voltage controlled stochasticity in performing low-voltage 8-bit analog to digital conversions. This eliminates the need for comparators, unlike the CMOS-based Flash Analog-to-Digital converters (ADC). This device allows for a simple and compact design of low-power approximate ADCs, which have become a quintessential element with the recent developments in neuromorphic computing and ``Internet of Things".  \end{abstract}
	
	\vspace{-3mm}

	%
	\IEEEpeerreviewmaketitle

	\section{Introduction}
	%
	%
	%
	%
	\IEEEPARstart{M}{agnetization} dynamics of nanomagnets is a function of temperature. In memory applications, reasonably high current is applied to devices based on nanomagnets to achieve deterministic switching across a range of temperature. Recently, however, the stochasticity introduced by the temperature dependence of nanomagnetic devices has been leveraged to implement several low-input applications such as biased random number generator\cite{Fukushima_2014}, Boolean and non-Boolean computations \cite{Sengupta_2016}, spiking neural networks\cite{sengup_stochastic_nn} and more recently, optimization based on Ising computations\cite{Shim_2017}. Recently, devices based on superparamagnetic magnets, with low energy-barrier ($E_B \sim 1kT$) between the two magnetic states, have been experimentally demonstrated to perform ultra-low power computations \cite{Mizrahi_2016} at the rate of tens of MHz\cite{vodenicarevic2017low} which can potentially go up to gigahertz range. Moreover, the probabilistic nature of these devices can be controlled by application of appropriate bias. Theoretical studies have also been performed on using low $E_B$ nanomagnets to mimic the Ising model and solve NP-complete optimization problems \cite{Sutton_2017}. Such controllable stochastic behavior also makes these devices suitable for data converters. Data converters have been traditionally based on CMOS \cite{Ginsburg_2007,El_Chammas_2010} but more recently spintronic devices have been explored to achieve lower power \cite{Yogendra_2015, Dong_2015}. Switching of magnetic tunnel junctions (MTJs) has also been explored to implement an ADC\cite{Choi_2015} but such implementations require additional circuitry for operation, such as external pulse generators. Note, however, that the devices explored are all current driven. The revival of multi-ferroic materials has enabled voltage-driven magnetization switching through the Magneto-electric (ME) effect \cite{Revival_ME}. Voltage driven devices based on ME effect promise to achieve lower switching energy\cite{Jaiswal_2017} than their current-driven counterparts. \par
	In this work, we propose a voltage controlled stochastic switching device based on a superparamagnetic nanomagnet which can perform approximate analog-to-digital (A-to-D) conversion at low voltages with up to 8-bit precision. By monitoring the states with a thresholding device such as a CMOS inverter, we show that the influence of the analog input voltage on the stochasticity of the device can naturally be digitally encoded. We leverage the $ns$-scale switching time to implement A-to-D conversions without the use of external pulse generators. The key highlights of the present work are: 1) We analyze the stochastic switching behavior, driven by magnetoelectric interaction in a voltage-controlled spintronic device comprising of low barrier magnets and elucidate the fundamental differences in its stochasticity from its high barrier counterpart. 2) We utilize the influence of analog voltage on the stochastic switching of the device to implement a low voltage ADC. The designed ADC promises to achieve compact and simple design as compared to state-of-the-art ADCs in contemporary technologies.
\vspace{-2mm}			
\section{\label{sec:level1}Lifetime of a Magnetic State}

\begin{figure}[t]
		\centering
		\includegraphics[width=0.5\textwidth]{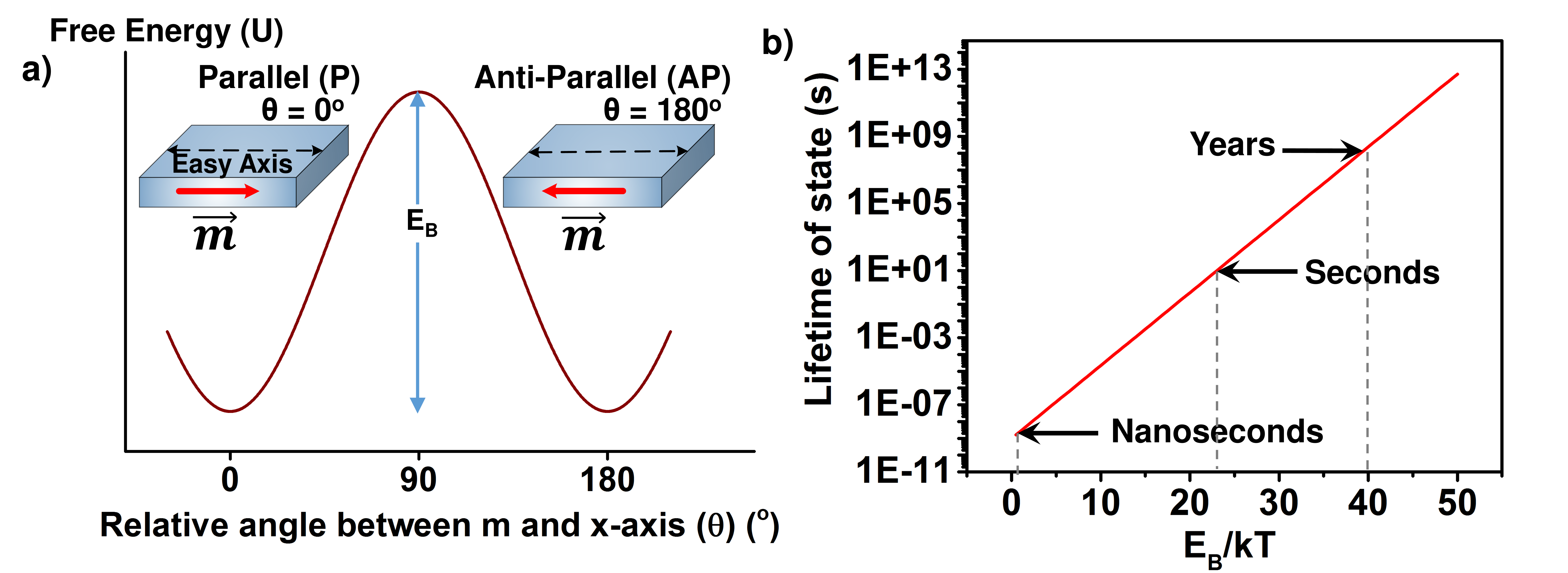}
		\caption{(a) Free energy profile of a nanomagnet showing two stable states, separated by a barrier $E_B$. Orientation of the magnet along easy axis corresponds to lowest energy. (b) Lifetime of the states as a function of $E_B$ showing switching times in the order of $ns$ for $\sim 1kT$ magnets\cite{Sun_2006}.}
		\vspace{-4mm}
		\label{fig:lifetime}
	\end{figure}

The energy profile of a nanomagnet shown in Fig. \ref{fig:lifetime}(b) gives us insight on its stability. The two low-energy states are separated by an energy barrier ($E_B$) which is essential for the nanomagnet to retain its alignment along the easy axis (x-axis). On receiving an input stimulus, $\vec{m}$ proceeds towards the opposite stable state. The magnitude of input stimulus determines how often the magnet switches between the two stable states for a given $E_B$. Interestingly, if this barrier is low enough, thermal energy alone may be enough to switch the state of the magnet. Fig. \ref{fig:lifetime}(c) shows the average lifetime of being in either of the two stable states ($t_l$), due to thermal energy alone. $t_{l}$ is thus expressed as\cite{Sun_2006} $t_{l} = t_{l0} exp(\frac{E_B}{kT})$ where $k$ is the Boltzmann constant and $T$ is the temperature. Typical values of $t_{l0}$ are between $0.1-1 ns$. High $E_B (\sim 40kT$) nanomagnets possess $t_l$ in the order of years which makes them useful in non-volatile memory and logic applications to ensure high retention times. On the other hand, for $E_B = 1 kT$, $t_l$ reduces to the order of nanoseconds (ns), and we observe stochastic switching between the states at gigahertz rates. This can be effected by mere thermal agitation, in the absence of an external stimulus. This concept of energy barrier is regularly used to distinguish between two states of the ferromagnetic layer with respect to another ferromagnetic layer with a fixed magnetization direction. MTJ is such a device, which we will discuss later. Note that the transition from ferromagnetism to superparamagnetism ($t_l \rightarrow t_{l0}$ in Eqn 1) has been explored in literature and the switching timescale was shown to be in good agreement with Eqn (1) \cite{Lopez_Diaz_2002}. Various applications have also been proposed\cite{Sutton_2017, Camsari_2017}, where Eqn (1) has been used to account for the switching times in the nanosecond order.
	\vspace{-2mm}  
	
	\section{A Voltage Controlled Stochastic Device}

	Ferromagnets are traditionally switched using the current induced spin transfer torque (STT) phenomenon \cite{Slonczewski_1996}. However, the STT phenomenon leads to sub optimal device performance due to various demerits, such as high write current requirement, switching asymmetry and shared read-write paths \cite{jaiswal_sca}. Some of these issues are mitigated using the spin hall effect (SHE) based ferromagnetic switching which lowers the switching current due to high spin-injection efficiency. However, recent advances in multi-ferroic materials have enabled low-energy voltage-induced ferromagnetic switching using magneto-electric (ME) effect \cite{Revival_ME}. Using ME effect, a transverse magnetic field is induced by an electric field, which is capable of switching the magnetization of the ferromagnet. A full electric field control of exchange bias and reversible switching of the ferromagnet was achieved using an MTJ stacked on a multi-ferroic layer, BiFeO$_3$  \cite{ramesh_me}. In this paper, we use the ME-effect to achieve voltage controlled switching in a nanomagnetic device.\par
 	\begin{figure}[]
		\centering
		\includegraphics[width=3.5in,keepaspectratio]{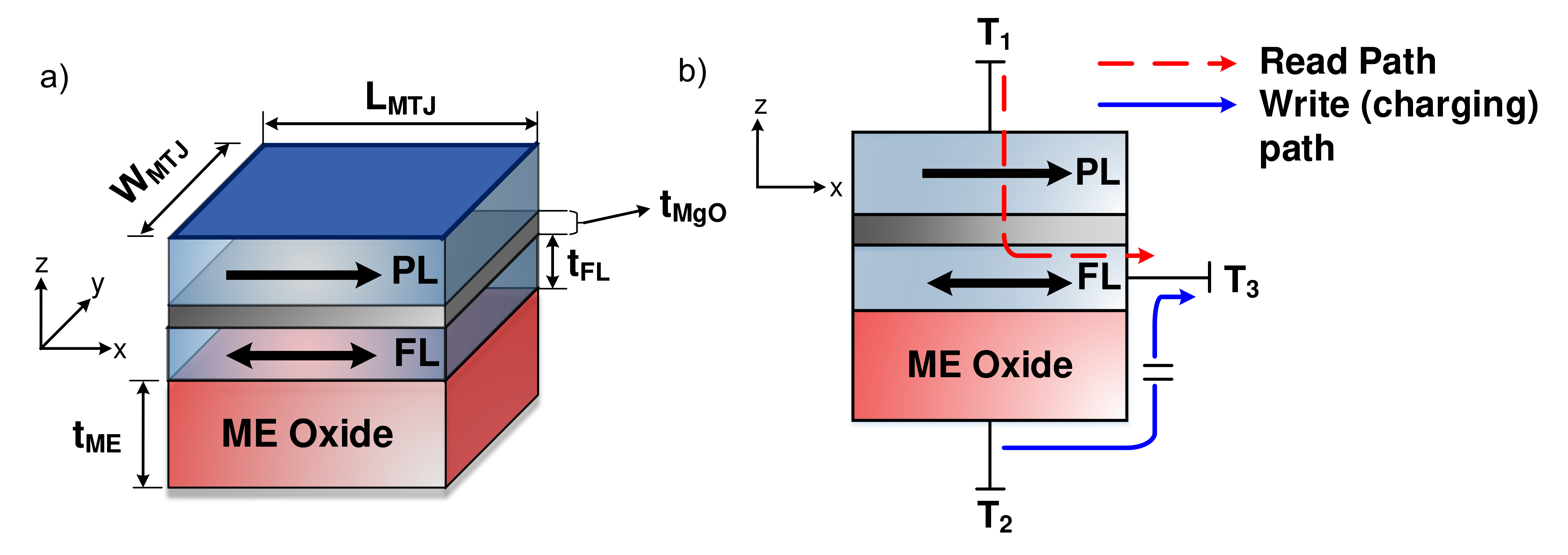}
		\caption{(a) Schematic of an ME-MTJ device where a MTJ is placed in contact with an ME oxide. (b) Schematic of a three terminal device is shown where the read and write paths are specified. Terminal T2 connects to the input voltage, while terminal T3 connects to the ground. Terminal T1 connects to the sensing circuit.}
		\vspace{-4mm}
		\label{fig:struc}
	\end{figure} 
	Fig. \ref{fig:struc} (a) shows the schematic of the device in concern, magneto-electric magnetic tunnel junction (ME-MTJ). The MTJ consists of two ferromagnetic layers - free layer (FL) and pinned layer (PL), separated by a thin tunneling oxide (usually MgO \cite{Butler_2001}). In the device, the MTJ is placed on top of an ME oxide layer. The magnetization of PL is fixed using an underlying anti-ferromagnetic (AFM) layer. The magnetization of the FL can be controlled by the ME oxide layer (BiFeO$_3$) in contact with the FL. The dimensions and other material parameters were chosen to control $E_B$. The two stable low-energy states of the ME-MTJ are the parallel state (P) and anti-parallel state (AP) according to the orientation of FL and PL. The input analog voltage is applied at terminal T2 of the device (refer Fig. \ref{fig:struc} (b)). The magnitude and polarity of this voltage control the magnetization of FL through ME effect, thus controlling the ME-MTJ switching behavior. The ME-MTJ resistance in P (AP) state is low (high). Thus, the switching behavior is observed by sensing the change in resistance of the ME-MTJ through the sensing circuit connected to terminal T1 of the device. Terminal T3 is grounded. The sensing circuit is described later in detail.
\vspace{-3mm}
	\subsection{Device Modeling}	
		
			
	The device described in the previous section is modeled using the Landau-Lifshitz-Gilbert (LLG) equation for magnetization dynamics, which in its implicit form is expressed as \cite{Gilbert_2004}: 
	
	\begin{equation}
    \vspace{-1mm}
	\frac{\partial \hat{m}}{\partial t}=-|\gamma|  \hat{m}\times H_{eff}+\alpha \hat{m}\times\frac{\partial  \hat{m}}{\partial t}
	\end{equation}
	where $\hat{m}$ is the unit magnetization vector, $\gamma$ is the gyromagnetic ratio, $\alpha$ is the Gilbert damping constant and $H_{eff}$ is the effective magnetic field. $H_{eff}$ is the sum of
	the demagnetization field, the interface anisotropy field \cite{Jaiswal_2016} and any other external field.  The models for demagnetization and the interfacial anistropy fields are detailed in existing works \cite{jaiswal2017energy}. The thermal noise is modeled using the Brown's model \cite{Brown_1963} and is accounted for by expressing a contributing field to $H_{eff}$ as: 
	\begin{equation}
    \vspace{-2mm}
	\vec{H}_{thermal} = \vec{\zeta} \sqrt {\frac{2\alpha kT}{|\gamma|M_SVdt}}
	\end{equation}	
	where $\vec{\zeta}$ is a vector with components that are standard-normal random variables, $V$ is volume of the free layer, $T$ is the temperature, $M_s$ is the saturation magnetization and $k$ is the Boltzmann's constant and $dt$ is the simulation time step. $E_B$ is expressed as $E_B=K_{u2}V$, where $K_{u2}$ is the second order uniaxial magnetic anisotropy constant.	The ME effect can be included in $H_{eff}$ by writing the ME field as \cite{manipatruni2015spin}
	\begin{equation}
    \vspace{-2mm}
	\vec{H}_{ME} = (\frac{1}{\mu_0}\alpha_{ME}\frac{V_{ME}}{t_{ME}}\hat{x}, 0\hat{y}, 0\hat{z})
	\end{equation}
	where the magneto-electric constant is $\alpha_{ME}$ \cite{Nikonov_2014}, $t_{ME}$ is the ME layer thickness and $V_{ME}$ is the voltage across the ME capacitor thus formed. Due to multifarious mechanisms proposed for possible voltage-driven switching \cite{ramesh_me,He_2010}, we have used a generic abstracted parameter $\alpha_{ME}$ for calculating $H_{ME}$. In theory, magnetoelectric effect can be expressed in terms of the equations\cite{eerenstein2006multiferroic} relating induced electrical polarization due to magnetic field and vice versa: $P_i = \alpha_{ij}H_j+(\beta_{ijk}/2)H_jH_k+...$ and $\mu_0M_i = \alpha_{ji}E_j+(\gamma_{ijk}/2)E_jE_k+...$ where $\alpha_{ij}$ and $\beta_{ijk}$ are second and third-rank tensors respectively. In practice, often the electric and magnetic fields resulting due to magneto-electric effect can be approximated \cite{lottermoser2004magnetic} by $P$ and $M$ respectively. The linear term $\alpha_{ij}$ dominates for materials with high permittivity and permeability. Measurements of such magneto-electric coupling often prove to be challenging and usually involve measuring the voltage representing the magnetization induced electric field as $\alpha = dE/dH$\cite{eerenstein2006multiferroic}, assuming the dominance of the linear term. Our aim here is to demonstrate the ADC operation, by sampling the telegraphic switching of the low-barrier magnet over a long duration. A non-linear $\alpha_{ME}$ would only affect the switching characteristics of the magnet, and not the stochastic switching behavior averaged over a long duration. Thus, a linear approximation of Eq. 4 is justified for demonstrating the ADC operation. Similar abstraction of ME parameters can be found in many previous works in literature \cite{Intel_ME,Ganguly_2016}.
\begin{figure}[t]
		\centering
		\includegraphics[width=0.5\textwidth,keepaspectratio]{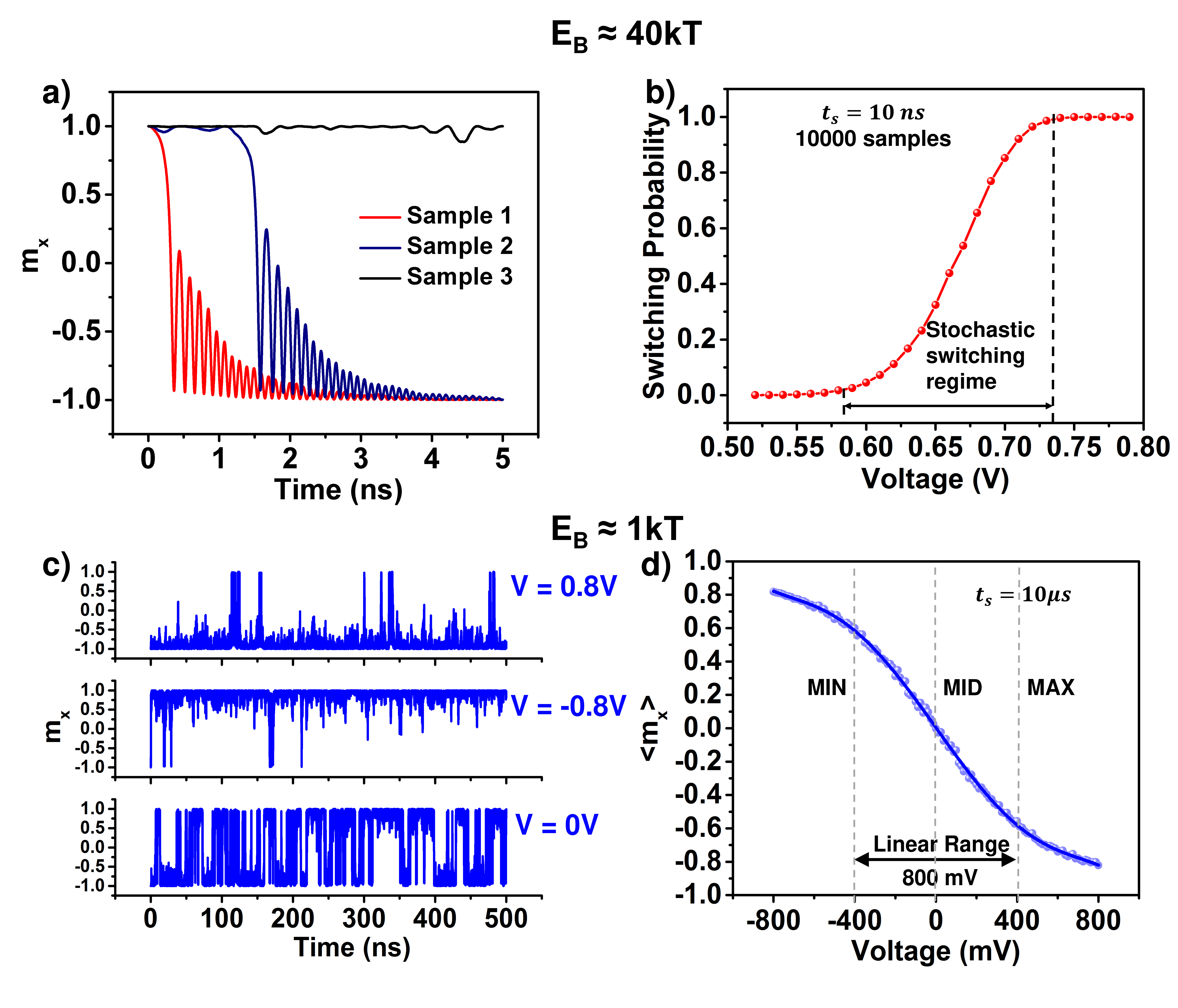}
		\vspace{-10mm}
		\caption{(a) Magnetization dynamics for device with $E_B =40kT$ showing variations for different sample experiments. (b) Switching probability curve v/s Voltage across 10000 sample experiments. (c) Magnetization dynamics over 500 ns for $V=0.8V$, $V=-0.8V$ and $V=0V$ for device with $E_B = 1kT$ demonstrating voltage control over stochastic switching. (d) $\langle m_x\rangle$, averaged over 10000 ns, decreases with voltage exhibiting a linear trend between -0.4V to 0.4V. 3 points namely MIN (-0.4V), MAX (0.4V) and MID (0V) are marked to denote the two extremities and the midpoint of the linear region.}
		\vspace{-5mm}
		\label{fig:swcurve}
	\end{figure}	
Eq. 2 can be solved numerically through the Heun's method \cite{Scholz_2001}. In addition, we used the non equilibrium Green's function (NEGF) formalism, verified with experimental benchmark, \cite{Fong_2011} for estimation of the resistance of the MTJ stack as a function of applied voltage and the magnetization directions. In this paper, we primarily focus on a low barrier nanomagnet with dimensions $20\times10\times1.35$ $nm^3$, while for high $E_B$ nanomagnet simulations, a device of dimensions $150\times60\times2.5$ $nm^3$ was used. Other relevant parameters are $t_{ME} = 5 nm, t_{MgO} = 1.8nm, M_s = 600.3 kA/m, \alpha = 0.012, K_{u2} = 15.3 kJ/m^3, K_i $(Interface anisotropy) $= 0.01 mJ/m^2, \alpha_{ME} = 0.05/c m/s, T = 300K$. Here, $c$ is the speed of light.

      \vspace{-3mm}
	\subsection{Influence of $E_B$ on Stochastic Switching}
 The role of thermal noise and external stimulus on the switching behavior fundamentally depends on $E_B$. For high $E_B$ devices, the main driving force is the magnetoelectric interaction and thermal noise acts as an instigator of stochasticity. The variation of magnetization switching dynamics for different runs, confirms the thermal stochasticity in this device, as shown in Fig. \ref{fig:swcurve} (a). The probability of switching increases with the magnitude of the input voltage as illustrated in Fig. \ref{fig:swcurve} (b). The contributions of thermal noise and the magnetoelectric effect are comparable in the stochastic switching regime shown.\par
	\begin{table}[b]
		\centering
          \vspace{-6mm}
		\caption{NRMSD for different bit precisions}
		\label{table:nrmsd}
		\begin{tabular}{|l|l|l|l|}
\hline
Bit Precision & 4-bit & 6-bit & 8-bit \\ \hline
NRMSD         & 3.8   & 2.71  & 2.11  \\ \hline
\end{tabular}
	\end{table}	
	On the other hand, for $E_B \sim 1 kT$, thermal noise can initiate probabilistic switching exclusively without the influence of external interaction. The applied voltage is therefore used to influence the stochasticity. As the lifetime of a particular state is in the order of $ns$, we choose the time-average of $m_x$, $\langle m_x\rangle$, over a sampling time, $t_s$ (say 10000ns), as a representative metric for the probability of the magnet being in a particular state. In Fig. \ref{fig:swcurve}(c) we plot $m_x$ over a 500ns snippet for the FL of the ME-MTJ, for input voltages $0.8V, -0.8V$ and $0V$ respectively. It is observed that in the absence of an external stimulus, the FL tends to switch back and forth uniformly over $t_s$, thus resulting in $\langle m_x\rangle \approx 0$ when averaged over a significant sample period. However, input stimulus manipulates the probability of the magnet being in a particular state as shown in Fig. \ref{fig:swcurve}(d). The accuracy of probability in Fig. \ref{fig:swcurve}(b) and Fig. \ref{fig:swcurve}(d) depends on the extent of sampling. For the low $E_B$ device, $t_s$ was chosen such that the variation in $m_x$ for a particular input across multiple experiments was minimized to $\pm 0.03$. We observe that the $\langle m_x\rangle$ decreases as the voltage varies from negative to positive. Our region of concern for ADC implementation is the linear range (shown in Fig. \ref{fig:swcurve}(d)) where it shows approximately linear behavior. It is to be noted that this linear range is a section of the sigmoid-like curve which is a characteristic curve of switching probability vs input stimulus for devices involving both high $E_B$ and low $E_B$ nanomagnets \cite{sengupta2016probabilistic, Camsari_2017}. The sigmoid behavior results from solving the stochastic LLG (Eq. 2) in presence of thermal noise \cite{Fukushima_2014}, which is modeled using a normal distribution (Eq. 3). Thus, the approximately linear operation of the ADC depends on the distribution of thermal fluctuations in nanomagnets. The normalized root mean square deviation (NRMSD) denotes the extent of deviation from the ideal linear behavior of $\langle m_x\rangle$ in the linear range. The number of voltage points (N) can be represented in terms of digital bit precision(m) such that $N = 2^m+1$ for relevance to the ADC we will propose in the next section. We calculated the NRMSD for 3 different bit precisions (Table. \ref{table:nrmsd}) which shows that the measurements deviate approximately 2-4 \% from ideal linear behavior. The deviations of the curve shown in Fig. \ref{fig:swcurve} (d) is dictated by $t_s$. Longer $t_s$ minimizes the deviation. Next, we describe how this behavior can be directly sensed and digitally encoded for data conversion applications.
    
\vspace{-3mm}
	\begin{figure}[t]
		\centering
		\includegraphics[width=3.4in,keepaspectratio]{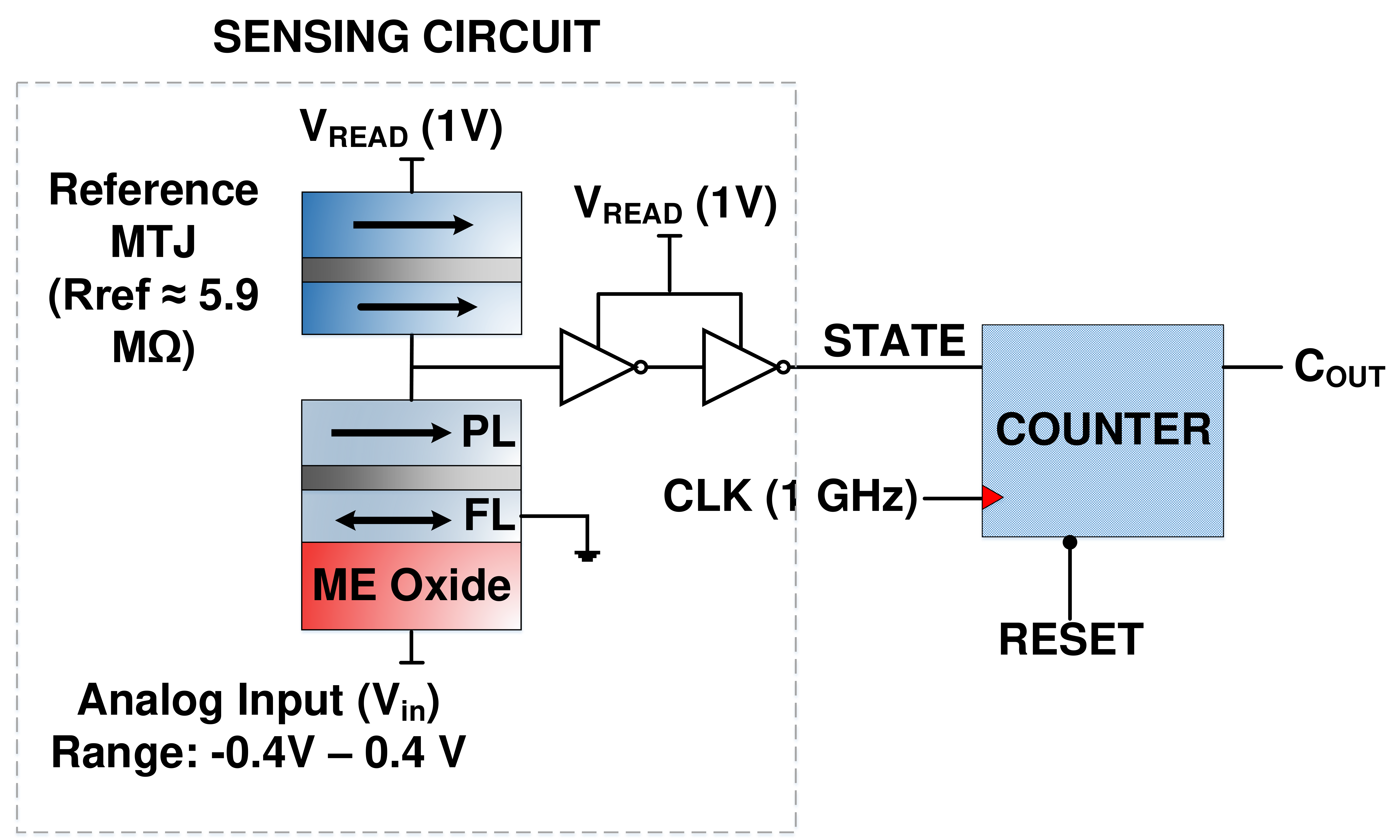}
		\caption{Circuit Schematic of the ADC implementation. Analog input is fed at the ME oxide terminal of the MTJ and digital output is obtained at the counter output.}
		
		\vspace{-4mm}
		\label{fig:ckt}
	\end{figure}

	
	\subsection{Device Readout}
	 The sensing circuit, shown in Fig. \ref{fig:ckt}, consists of a reference MTJ ($R_{Ref}$) forms a voltage divider with the resistance of the MTJ ($R_{MTJ}$). The resistance in the P state ($R_P$) is lower than the resistance in the AP state ($R_{AP}$). 
     Considering the magnetization of PL ($m_{x,p}$) pinned along x-axis, $R_{MTJ} = R_{AP} (R_P)$ when $m_x \approx -1 (+1)$. The $t_{MgO}$ of the MTJs are optimized such that the read current is in the order of $\sim 50 nA$ to ensure minimum influence on the switching probability curve shown in Fig. \ref{fig:swcurve}(b). The voltage divider output is sensed using two back to back inverters. The chain of inverters ensure a rail-to-rail swing at the inverter output ($STATE$ in Fig. \ref{fig:ckt}).
     \vspace{-2mm}	
	\section{ADC Implementation}
	
	\label{sec:ADCimpl}
\begin{figure}[t]
		\centering
		\includegraphics[width=3.6in, keepaspectratio]{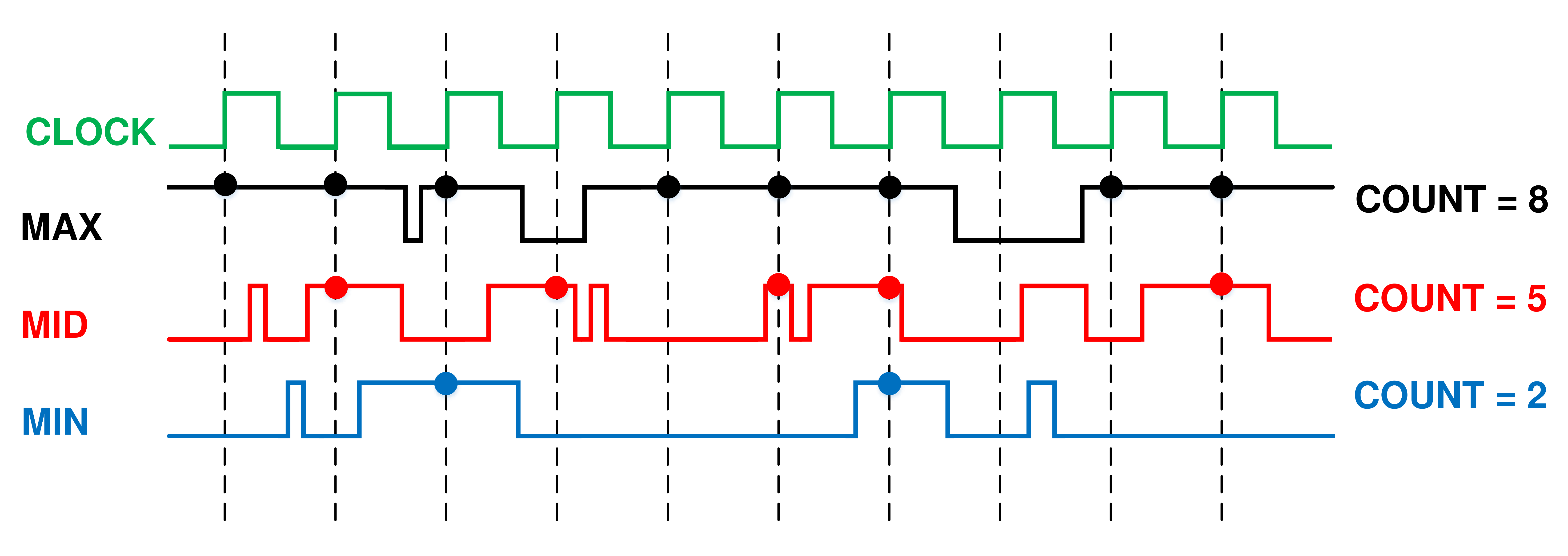}
		\caption{Illustration showing how the digital output is obtained from the inverter output ($STATE$). 3 sample cases for $STATE$ is shown. The clock samples the inverter output at each positive edge. If output is `1' counter counts up (shown by dots) thus making `COUNT' proportional to the `STATE' variable.}
		\label{fig:countcartoon}
	\end{figure}

	The influence of an analog voltage on the stochastic switching of the spintronic device, described earlier, was used to implement a low-voltage 8-bit ADC. An increase in the input voltage decreases $\langle m_x \rangle$, thus increasing the probability of the magnet being in AP state. When $R_{MTJ} = R_{AP} (R_P)$ or $m_x \approx -1 (+1)$, the input to the inverter chain is high (low), hence the $STATE=1 (0)$. Thus a low $m_x$ indicates a high $STATE$. $STATE$ is further fed to the data input of a ripple counter, supplied by a $1~GHz$ clock. The counter counts up each time the value of $STATE$ is logic `1', at the positive edge of the clock. This is illustrated in Fig. \ref{fig:countcartoon}. The digital count thus obtained can be translated to a binary code using look-up tables \cite{Choi_2015}.

	\begin{figure}[t]
		\centering
		\includegraphics[width=3.6in,keepaspectratio]{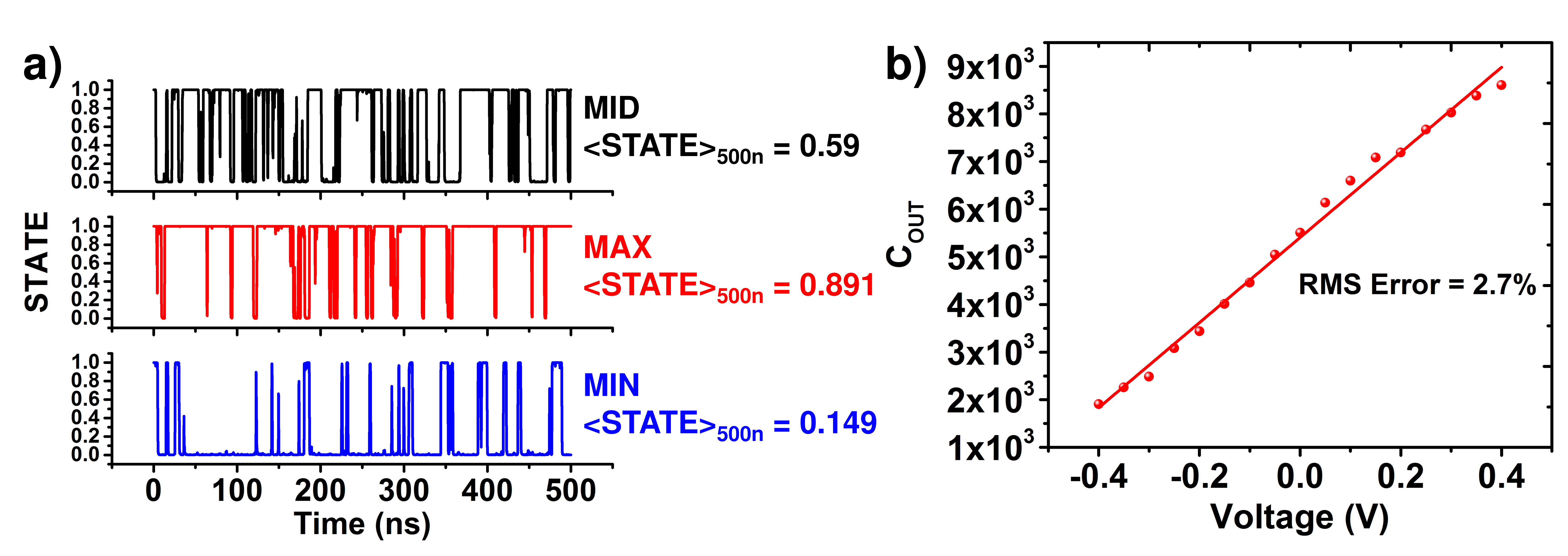}
		\caption{(a) The input to the counter is plotted as a function of time for 500 ns for 3 points MIN, MAX and MID as described earlier. $STATE$ thus reflects the magnetization characteristics of Fig. \ref{fig:swcurve}. (b) $C_{OUT}$ v/s Voltage shows linear trend desirable for ADC operation.}
		\vspace{-4mm}
		\label{fig:invout}
	\end{figure}
	The circuit described in Fig. \ref{fig:ckt} is simulated in HSPICE assuming a 45nm PTM technology\cite{PTM}. The modeling of the MTJ as a resistance is described earlier. To be noted, the ground terminal is connected to FL, hence the resistance of the ME oxide is not considered. As $m_x$ of FL, and hence $\theta$, varies over time, the value of $STATE$ is either logic `0' or logic `1' (approximately) at each time instant, as can be seen from Fig. \ref{fig:invout}(a). Fig. \ref{fig:invout}(a) also shows that at MAX, $STATE$ prefers `1', while at MIN, it prefers `0'. At MID however, value of $STATE$ is uniformly distributed between `1' and `0'. Thus, STATE reflects the magnetization characteristics shown in Fig. \ref{fig:swcurve}. By counting up each time $STATE$=`1' at the positive edge of the clock, $C_{OUT}$ essentially produces the output by sampling $STATE$ every 1 $ns$. Hence, $C_{OUT}$ is inversely proportional to the probability of the nanomagnet being in a particular state. The NRMSD for the linear trend of $C_{OUT}$ is 2.7\% for 17 points, which is reflective of the NRMSD obtained from the linearity trend of the magnetization curve in Table.\ref{table:nrmsd}. Thus, For $t_s = 10 \mu s$, an 8-bit precision was achieved with NRMSD of approximately 2 \%. Bit precision can further be increased with longer $t_s$.\par
	An 8-bit precision was achieved with an NRMSD error of $\approx $ 2\% for a sampling time of 10 $\mu s$. The worst case power consumption during sensing and counting is $34 \mu W$. The energy consumption can be reduced by accommodating for a lower accuracy or lower bit precision. The input range of the ADC is -0.4V to 0.4V which makes it suitable for sensors that require low-voltage conversions. It is worth noting that high $E_B$ devices possess a higher switching voltage and lower switching speed than a low $E_B$ device. Moreover, since $E_B = K_{u2}V$, low $E_B$ magnets are area-eficient. As a result, circuits that utilize low $E_B$ devices promise to achieve lower power and compact area. 
\vspace{-2mm}	
\section{Conclusion}
	We explore a voltage controlled stochastic spintronic device based on superparamagnetic nanomagnets as a possible candidate to perform approximate low voltage Analog-to-Digital conversions.  The simulated device achieves voltage controllable telegraphic switching at a $ns$ scale which enables digital encoding of analog inputs without any modulation of the input. Moreover, the low voltage operation allows for obvious power benefits. Thus, the design paves the way for a compact ADC, which is a quintessential element in light of recent developments in the field of neuromorphic computing and ``Internet of Things".  
\section*{Acknowledgements}
The work was supported in part by, C-SPIN, a MARCO and DARPA sponsored StarNet center, by the Semiconductor Research Corporation, the National Science Foundation, Intel Corporation and by the DoD Vannevar Bush Fellowship.
\vspace{-4mm}	
	
	\bibliographystyle{IEEEtran}
	\bibliography{ADC.bib}

\end{document}